\documentclass[12pt]{article}
\usepackage{amsmath,epsfig}

\def\ignore#1{{}}

\newcommand{\adsc}{\bt}

\newcommand{\alp}{\alpha}
\newcommand{\bt}{\beta}
\newcommand{\gm}{\gamma}
\newcommand{\Gm}{\Gamma}
\newcommand{\dlt}{\delta}

\newcommand{\tht}{\theta}

\newcommand{\vtht}{\vartheta}
\newcommand{\kp}{\kappa}

\newcommand{\sgm}{\sigma}
\newcommand{\Sgm}{\Sigma}
\newcommand{\vph}{\varphi}
\newcommand{\omg}{\omega}
\newcommand{\Omg}{\Omega}

\newcommand{\vth}{\vartheta}

\newcommand{\be}{\begin{equation}}
\newcommand{\ee}{\end{equation}}
\newcommand{\bea}{\begin{eqnarray}}
\newcommand{\eea}{\end{eqnarray}}
\newcommand{\eql}{\!\!\!&=\!\!\!&}

\newcommand{\defa}{\!\!\!&\equiv\!\!\!&}

\newcommand{\simgt}{\stackrel{>}{{}_\sim}}

\newcommand{\tl}[1]{\tilde{#1}}
\newcommand{\bdm}[1]{{\mbox{\boldmath $#1$}}}

\newcommand{\der}{\partial}
\newcommand{\dr}{\!\!d}
\newcommand{\hc}{{\rm h.c.}}
\newcommand{\ie}{i.e.}

\newcommand{\vev}[1]{\langle #1 \rangle}

\newcommand{\brkt}[1]{\left( #1 \right)}
\newcommand{\brc}[1]{\left\{ #1 \right\}}
\newcommand{\sbk}[1]{\left[ #1 \right]}
\newcommand{\abs}[1]{\left| #1 \right|}

\renewcommand{\Re}{{\rm Re}\,}
\renewcommand{\Im}{{\rm Im}\,}

\newcommand{\cA}{{\cal A}}

\newcommand{\cF}{{\cal F}}

\newcommand{\cH}{{\cal H}}

\newcommand{\cL}{{\cal L}}

\newcommand{\cN}{{\cal N}}
\newcommand{\cO}{{\cal O}}

\newcommand{\cV}{{\cal V}}

\renewcommand{\ge}[2]{e_{#1}^{\;\;#2}}

\newcommand{\nV}{n_V}
\newcommand{\nVp}{n'_V}
\newcommand{\nH}{n_H}

\newcommand{\gey}{\ge{y}{4}}
\newcommand{\SUu}{SU(2)_{\mbox{\scriptsize $\bdm{U}$}}}
\newcommand{\thU}{\theta_{\mbox{\scriptsize $\bdm{U}$}}}

\numberwithin{equation}{section}
\renewcommand{\thefootnote}{\fnsymbol{footnote}}

\begin{document}

\title{
\begin{flushright}
\ \\*[-80pt] 
\begin{minipage}{0.25\linewidth}
\normalsize
arXiv:0709.3791 \\
YITP-07-57 \\
OU-HET 586/2007 
\\*[50pt]
\end{minipage}
\end{flushright}
{\Large \bf 
Moduli stabilization in 5D gauged supergravity \\
with universal hypermultiplet and boundary 
superpotentials
\\*[20pt]}}

\author{Hiroyuki~Abe$^{1,}$\footnote{
E-mail address: abe@yukawa.kyoto-u.ac.jp} \ and \ 
Yutaka~Sakamura$^{2,}$\footnote{
E-mail address: sakamura@het.phys.sci.osaka-u.ac.jp} \\*[20pt]
$^1${\it \normalsize 
Yukawa Institute for Theoretical Physics, Kyoto University, 
Kyoto 606-8502, Japan} \\
$^2${\it \normalsize 
Department of Physics, Osaka University, 
Toyonaka, Osaka 560-0043, Japan} \\*[50pt]}

\date{
\centerline{\small \bf Abstract}
\begin{minipage}{0.9\linewidth}
\medskip 
\medskip 
\small
We study a four-dimensional effective theory of the 
five-dimensional (5D) gauged supergravity with a universal 
hypermultiplet and perturbative superpotential terms at the orbifold 
fixed points. 
Among eight independent isometries of the scalar manifold, 
we focus on three directions for gauging by the graviphoton. 
The class of models we consider includes the 5D heterotic M-theory 
and the supersymmetric Randall-Sundrum model 
as special limits of the gauging parameters. 
We analyze the vacuum structure of such models, 
especially the nature of moduli stabilization, 
from the viewpoint of the effective theory. 
We also discuss the uplifting of supersymmetric anti-de Sitter vacua. 
\end{minipage}
}

\begin{titlepage}
\maketitle
\thispagestyle{empty}
\end{titlepage}

\ignore{
\clearpage
\thispagestyle{empty}
\tableofcontents
\clearpage
}

\renewcommand{\thefootnote}{\arabic{footnote}}
\setcounter{footnote}{0}
\setcounter{page}{1}

\section{Introduction}
Supersymmetric (SUSY) extension of 
the Randall-Sundrum (RS) model~\cite{Randall:1999ee,
Altendorfer:2000rr,Gherghetta:2000qt} provides an interesting setup 
for the physics beyond the standard model. 
For instance, localized wavefunctions in the extra dimension can be 
considered as a source of Yukawa hierarchy~\cite{ArkaniHamed:1999dc} 
as well as the large hierarchy between the weak 
and the Planck scales. 
The SUSY breaking 
sector can be sequestered from the visible sector in the 
extra dimension resulting in a flavor blind SUSY breaking 
patterns of the superparticle masses and couplings, 
i.e., anomaly mediation~\cite{Randall:1998uk}. 
Besides, the application of the so-called AdS/CFT 
correspondence~\cite{Maldacena:1997re} to the SUSY RS model 
might provide a way to analyze perturbatively the 
four-dimensional (4D) strongly coupled theories. 

On the other hand, the superstring theories would provide 
a unified framework of the standard model and gravity. 
It is known that a low energy effective theory of 
the strongly coupled heterotic string theory~\cite{Horava:1995qa} 
can be described by 5D supergravity on an 
orbifold, which is sometimes referred to as 
the 5D heterotic M-theory~\cite{Lukas:1998yy}. 
Some three-brane solutions have been derived in such 5D model, 
which can be utilized to construct brane-world scenarios 
of our universe, where the visible sector and the hidden 
sector reside in different branes located at different 
orbifold fixed points. 

All the above examples can be categorized into a unique 
theory, i.e., the 5D {\it gauged} supergravity, and share a common 
important subject, that is, an issue of moduli stabilization. 
5D models have at least a radius modulus, \ie, {\it the radion}, 
whose vacuum value corresponds to the radius of the extra dimension. 
In 4D effective theory of 5D models, some couplings 
depend on the radius which is undetermined unless 
the radion is dynamically stabilized. 
If a 5D model is an effective theory of some 
higher-dimensional model, there might exist other moduli. 
These moduli fields including the 
radion form supermultiplets in the 4D effective 
theory if the compactification respects $N=1$ SUSY, and 
these multiplets can mediate SUSY breaking effects. 
Thus the moduli stabilization is one of the most 
important issues in the model building based on 
the higher-dimensional theory. 

Several years ago, an interesting class of 
the moduli stabilization scheme was proposed by 
Kachru-Kallosh-Linde-Trivedi (KKLT)~\cite{Kachru:2003aw} 
based on the type IIB supergravity~\cite{Giddings:2001yu}. 
In such framework, light moduli are first stabilized 
at a SUSY preserving anti-de Sitter (AdS) minimum of 
the scalar potential. Then it is uplifted to a Minkowski 
minimum by a SUSY breaking vacuum energy generated in 
the hidden sector such as anti D-branes~\cite{Kachru:2003aw,Choi:2004sx}, 
or dynamically generated F-terms~\cite{Saltman:2004sn,Dudas:2006gr} 
and D-terms~\cite{Burgess:2003ic}, 
which are well sequestered from the light moduli 
as well as the visible sector. 
This scenario overcomes, in a controllable manner, 
a big difficulty existing in supergravity/string 
compactifications~\cite{Maldacena:2000mw}, 
that is, a realization of the SUSY breaking 
Minkowski vacuum where all the moduli are stabilized, 
which is required from the observation of our universe. 

Moreover, it has been shown that this kind of moduli 
stabilization procedures generically yields an 
interesting pattern of soft SUSY breaking terms 
in the visible sector, that is, the mirage mediation~\cite{
Choi:2004sx,Choi:2005uz,Endo:2005uy}. 
In the mirage mediation, the modulus-mediated contribution 
is comparable to that of the anomaly mediation. 
The low energy superparticle spectrum 
is quite different from the other mediation schemes 
such as the pure modulus mediation, the pure anomaly 
mediation and the gauge mediations. These would be 
distinguished by high-energy experiments and 
cosmological observations in future. 

In this paper we study 5D gauged supergravity with a 
universal hypermultiplet whose isometry group is $SU(2,1)$. 
We consider various gaugings of the isometries and introduce 
some superpotential terms at the orbifold fixed points. 
Among eight independent isometries, we focus on three directions 
to gauge by the graviphoton. 
Thus the model we consider has three gauging 
parameters~$(\alp,\bt,\gm)$ and includes both the SUSY RS model 
with an arbitrary bulk mass parameter for the hypermultiplet 
and 5D heterotic M-theory as different choices of the parameters. 
We investigate the vacuum structure of such models, 
especially the nature of moduli stabilization, 
assuming perturbative superpotential terms at the fixed points. 
Motivated by the KKLT moduli stabilization scheme, we also 
discuss the uplifting of SUSY AdS vacua in our models 
and the resultant SUSY breaking. 

The following sections are arranged as follows. 
In Sec.~\ref{sec:effective}, we review the $N=1$ 
off-shell description of 5D supergravity and 
see the representation of the isometries on 
the scalar manifold in such description. 
Then we derive the 4D effective action of 
a class of models obtained by gauging three 
independent isometries by the graviphoton 
with arbitrary superpotential terms at the 
orbifold fixed points. 
In Sec.~\ref{sec:hybrid}, we study the moduli stabilization and 
the uplifting of the scalar potential 
in a case where the effective K\"{a}hler and 
superpotentials are expressed by analytic functions. 
In Sec.~\ref{sec:gls}, 
we carry out a similar analysis in a model 
which corresponds to the generalized 
Luty-Sundrum model~\cite{Luty:2000ec}. 
Sec.~\ref{sec:summary} is devoted to the summary. 
In Appendix, we show how to derive the isometry transformations 
in the on-shell description of 5D supergravity 
from the off-shell formulation.

\section{4D effective action of 5D gauged supergravity}
\label{sec:effective}

\subsection{N=1 off-shell description of 5D supergravity action}
In this paper we consider 5D gauged supergravity compactified 
on an orbifold~$S^1/Z_2$ with a universal hypermultiplet and arbitrary 
superpotentials at the fixed points (or the boundaries) 
of the orbifold. 
The metric is written as 
\be
 ds_5^2 = e^{2\sgm(y)}g_{\mu\nu}dx^\mu dx^\nu-(\gey dy)^2, 
\ee
where $\sgm(y)$ is a warp factor. 
The off-diagonal components of the metric~$g_{\mu y}$ is gauged away. 
We take the fundamental region of the orbifold as $0\leq y\leq \pi R$, 
where $R$ is a constant,\footnote{
In principle, $R$ is nothing to do with the radius of the orbifold. 
It coincides with the latter only when the coordinate $y$ is redefined 
so that $\vev{\gey}=1$.}
and take the unit of $M_5=1$, where $M_5$ is the 5D Planck mass. 

\ignore{
In this section, we derive a 4D effective theory of the 5D 
gauged supergravity with a universal hypermultiplet 
and arbitrary superpotentials at the orbifold fixed points 
as well as generic gaugings. Usual on-shell dimensional 
reduction is not useful for such purpose, because it is highly 
dependent to the explicit form of the superpotential and the 
gaugings. 
}

The off-shell description of 5D supergravity is quite useful for our purpose. 
It enables us to treat the localized terms at the orbifold boundaries 
independently from the bulk action. 
Furthermore as will be seen in the next subsection, 
the isometries of the scalar manifold, some of which are to be gauged, 
are {\it linearly} realized in the off-shell description. 
Therefore we start from 5D off-shell (conformal) supergravity 
developed by Ref.~\cite{Kugo:2000af}-\cite{Kugo:2002js}. 
5D superconformal multiplets relevant to our study 
are the Weyl multiplet $\mbox{\boldmath $E$}_W$, 
the vector multiplets $\mbox{\boldmath $V$}^I$ and 
the hypermultiplets $\mbox{\boldmath ${\cal H}$}^{\hat{a}}$, 
where $I=0,1,2,\ldots,n_V$ and 
$\hat{a}=1,2,\ldots,n_C+n_H$. 
Here $n_C$, $n_H$ are the numbers of compensator and physical 
hypermultiplets, respectively. 
These 5D multiplets are decomposed into $N=1$ superconformal 
multiplets~\cite{Kugo:2002js} as $\mbox{\boldmath $E$}_W=(E_W,V_E)$, 
$\mbox{\boldmath $V$}^I=(V^I,\Sigma^I)$ and 
$\mbox{\boldmath ${\cal H}$}^{\hat{a}}
=(\Phi^{2 \hat{a}-1},\Phi^{2\hat{a}})$, 
where $E_W$ is the $N=1$ Weyl multiplet, $V_E$ is the $N=1$ general 
multiplet whose scalar component is $\gey$, 
$V^I$ is the $N=1$ vector multiplet, and $\Sigma^I$, 
$\Phi^{2 \hat{a}-1}$, $\Phi^{2 \hat{a}}$ are $N=1$ 
chiral multiplets. 

The 5D supergravity action can be written in terms of these $N=1$ 
multiplets~\cite{Paccetti Correia:2004ri}, in which 
$V_E$ has no kinetic term. 
After integrating $V_E$ out, the 5D action is expressed 
as~\cite{Correia:2006pj,Abe:2006eg} 
\begin{eqnarray}
{\cal L} &=& 
-3e^{2\sigma} \int d^4\tht\,\cN^{1/3}(\cV) 
\brc{d_a^{\ b} \bar\Phi^b (e^{-2igt_IV^I})^a_{\ c} \Phi^c}^{2/3} 
\nonumber \\ &&
-e^{3\sigma} \bigg[ 
\int d^2 \theta\, \Phi^a d_a^{\ b} \rho_{bc} 
(\partial_y-2igt_I\Sigma^I)^c_{\ d} \Phi^d 
+\textrm{h.c.} \bigg] \nonumber\\ && 
+\sum_{\vtht=0,\pi}\cL^{\vtht}\,\dlt(y-\vtht R)+\cdots, 
\label{eq:lbulk}
\end{eqnarray}
where $e^{2\sigma}$ is the warp factor of the background 
metric to be determined on-shell, 
$d_a^{\ b}={\rm diag}({\bf 1}_{2n_C},-{\bf 1}_{2n_H})$, 
$\rho_{ab}=i \sigma_2 \otimes {\bf 1}_{n_C+n_H}$ and 
$\cN(\cV)$ is a cubic function defined by 
\be
 \cN(\cV) \equiv C_{IJK}\cV^I\cV^J\cV^K,  \label{def_cN}
\ee
where $C_{IJK}$ is a real constant tensor which is completely symmetric
for indices, and $\cV^I\equiv -\der_y V^I+\Sgm^I+\bar{\Sgm}^I$ 
is a gauge-invariant quantity. 
The ellipsis in (\ref{eq:lbulk}) denotes the vector multiplet part. 
The boundary Lagrangian~$\cL^\vtht$ can be introduced 
independently of the bulk action. 
Note that (\ref{eq:lbulk}) is a shorthand expression for 
the full supergravity action. 
We can always restore the full action by promoting 
the $d^4\tht$ and $d^2\tht$ 
integrals to the $D$- and $F$-term formulae 
of $N=1$ conformal supergravity formulation~\cite{4Doffshell}, 
which are compactly listed in Appendix~C of Ref.~\cite{Kugo:2002js}. 
In the above expression of the 5D action, 
the hypermultiplet isometries are linearly realized. 
We can partially or fully gauge these isometries by 
the vector multiplets~$\bdm{V}^I=(V^I,\Sgm^I)$ 
with the generator $igt_I$. 

\ignore{
For instance, in the case $n_C=2$, that is, two compensator 
hypermultiplets\footnote{
One of them, $\mbox{\boldmath ${\cal H}$}^1$, should be 
eliminated by introducing and integrating out nondynamical 
vector multiplet. See, e.g., Refs.~\cite{Abe:2006eg} for detail.} 
$\mbox{\boldmath ${\cal H}$}^1=(\Phi^1,\Phi^2)$ and 
$\mbox{\boldmath ${\cal H}$}^2=(\Phi^3,\Phi^4)$ with 
the $Z_2$-parity assignment 
$\Pi(\Phi^1,\Phi^2,\Phi^3,\Phi^4)=(-,+,+,-)$, 
the boundary Lagrangian terms at the $y= \vartheta R$ 
fixed point are written as 
\begin{eqnarray}
{\cal L}^{\vartheta} 
&=& e^{3\sigma} \bigg[ \int d^2 \theta\, 
(\Phi^2 \Phi^3) P_\vartheta(Q) +\textrm{h.c.} \bigg] 
+\cdots, 
\label{eq:lbrane}
\end{eqnarray}
where the ellipsis stands for the terms including 
the K\"ahler potential and the gauge kinetic functions. 
If chiral multiplets represented by $Q$ appearing in 
the superpotential originate form bulk hypermultiplets, 
they are determined as $Q=\Phi^{2 \hat{a}-1}/\Phi_3$ in 
this example.  The compensator dependence in the boundary 
Lagrangian as well as in $Q$ is determined by the Weyl 
weight counting and so on~\cite{Abe:2006eg}. 
By setting all these data, the 5D action is completely 
determined. As next tasks, we usually fix the superconformal 
symmetries, integrate out all the auxiliary fields and obtain 
the corresponding on-shell 5D action. However, in the off-shell 
dimensional reduction procedure~\cite{Correia:2006pj,Abe:2006eg}, 
we postpone the superconformal gauge fixing until arriving 
at the 4D (superconformal) action. Instead of that, we first 
integrate out all the $Z_2$-odd $N=1$ multiplets by neglecting 
their kinetic terms. This can be performed at the superfield 
level, without violating the $N=1$ off-shell structure. 
Interestingly enough, this integration also subtract the 
zero-modes of the $Z_2$-even multiplets, which makes the 
derivation of the 4D effective theory easier. }

\subsection{Gauged supergravity with universal hypermultiplet}
Now we consider the gauged supergravity with a single universal 
hypermultiplet spanning the manifold $SU(2,1)/SU(2) \times U(1)$. 
The universal hyperscalars are commonly denoted as $S$ and $\xi$, 
which are even and odd under the orbifold parity 
\cite{attractor,Falkowski:2000yq}. 
The scalar manifold has an $SU(2,1)$ isometry group, 
which is nonlinearly realized in the on-shell description~\cite{attractor}. 
(See (\ref{trf:S-xi}) in Appendix~\ref{isometry}, for example.) 
In the off-shell description~(\ref{eq:lbulk}), on the other hand, 
it is {\it linearly} realized. 
This greatly simplifies the analysis. 
The situation we consider is realized by taking $(n_C,n_H)=(2,1)$.  
Then the bulk action in (\ref{eq:lbulk}) has a $U(2,1)$ symmetry. 
Since the superconformal gauge-fixing conditions can eliminate only 
one compensator multiplet, we introduce a nondynamical (auxiliary) 
Abelian vector multiplet~$\bdm{V}_T=(V_T,\Sgm_T)$ to eliminate 
the other compensator multiplet. 
We gauge the overall $U(1)$ subgroup of the symmetry group~$U(2,1)$, 
which we refer to as $U(1)_T$, by $\bdm{V}_T$~\cite{Fujita:2001bd}. 
Namely the charges of the hypermultiplets for this gauging are 
assigned as $igt_T=\sgm_3\otimes\bdm{1}_{2+n_H}$. 
As a result the symmetry group is reduced to $SU(2,1)$ 
after eliminating the nondynamical vector multiplet~$\bdm{V}_T$. 
The $Z_2$-parities and the $U(1)_T$ charges of the $N=1$ multiplets 
are listed in Table.~\ref{Z2_parity}. 
\begin{table}[t]
\begin{center}
\begin{tabular}{|c|c|c|c|c|c|c|c|c|c|c|c|c|} \hline
\rule[-2mm]{0mm}{7mm} & $V^{I'}$ & $\Sgm^{I'}$ & $V^{I''}$ & $\Sgm^{I''}$ 
 & $V_T$ & $\Sgm_T$ & $\Phi^1$ & 
 $\Phi^2$ & $\Phi^3$ & $\Phi^4$ & $\Phi^5$ & $\Phi^6$
 \\ \hline 
$Z_2$-parity & $-$ & $+$ & $+$ & $-$ & $+$ & $-$ & $-$ & $+$ & $+$ 
 & $-$ & $+$ & $-$ \\ \hline
$U(1)_T$ charge & 0 & 0 & 0 & 0 & 0 & 0 & 1 & $-1$ & 1 & $-1$ 
& 1 & $-1$ \\ \hline
\end{tabular}
\end{center}
\caption{The orbifold parities and the $U(1)_T$ charges of $N=1$ multiplet. 
The indices run over $I'=0,1,\cdots,\nVp$; $I''=\nVp+1,\nVp+2,\cdots,\nV$;
$\hat{a}=2,3,\cdots,\nH+2$. }
\label{Z2_parity}
\end{table}
Two compensator multiplets~$\bdm{\cH}^1=(\Phi^1,\Phi^2)$ and 
$\bdm{\cH}^2=(\Phi^3,\Phi^4)$ must have the opposite $Z_2$-parities 
for consistency with the superconformal gauge-fixing. 
(See the appendix in Ref.~\cite{Fujita:2001bd}.)
For the vector multiplets, we divide the index~$I$ into $(I',I'')$ 
so that $V^{I'}$ and $V^{I''}$ are odd and even 
under the $Z_2$-parity respectively. 

The following analyses are independent of the number of 
$Z_2$-even vector multiplets $n''_V$, and then we just choose 
$n''_V=0$ to simplify the discussion (\ie, $I=I'$). 
The gauged supergravity is obtained by gauging some directions within 
the isometry group~$SU(2,1)$ by $\bdm{V}^{I'}$. 
In this paper we restrict ourselves to the simple case $n'_V=0$, where 
only the graviphoton takes part in the gauging of the isometries. 
Namely the function~$\cN$ defined in (\ref{def_cN}) is now
\be
 \cN(\cV) = \brkt{\cV^0}^3. 
\ee
The most general form of the gauging is parameterized by 
\begin{eqnarray}
igt_{I=0} &=& \sum_{i=1}^8 \tilde\alpha_i \,T^i 
\label{eq:u21gg}
\end{eqnarray}
acting on $(\Phi^1,\Phi^3,\Phi^5)^t$ or $(\Phi^2,\Phi^4,\Phi^6)^t$, 
where $T^i$ ($i=1,\ldots,8$) are $3 \times 3$ matrix-valued 
generators of $SU(2,1)$ shown in Eq.~(\ref{eq:su21gen}). 
Here the real coefficients $\tilde\alpha_i$ determine the 
gauging direction. 
Since the graviphoton is $Z_2$-odd, 
the parameters~$\tl{\alp}_i$ ($i=1,2,4,5$) are $Z_2$-even while 
the others are $Z_2$-odd and have kink profiles for $y$.\footnote{ 
The $Z_2$-odd gauge couplings can be consistently introduced 
into supergravity by the so-called four-form mechanism proposed 
in Ref.\cite{BKV}. }
In the following, we consider a case that $\tl{\alp}_i$ are 
parameterized by three parameters~($\alp,\bt,\gm$) as  
\bea
\tilde\alpha_3 \eql 2 \beta, \qquad 
\tilde\alpha_6 = \alpha, \qquad 
\tilde\alpha_8 = \alpha+\beta+\gamma, \nonumber\\
\tl{\alp}_i \eql 0. \;\;\;\;\; (i\neq 3,6,8)
\eea

This class of models contain the following interesting models 
as special limits of the gauge parameters. 
In the limit of $\bt,\gm\to 0$, this model is reduced 
to the 5D effective theory of heterotic M-theory, which is 
derived in Ref.~\cite{Lukas:1998yy} on-shell. 
\ignore{
This is extended to more general gauging
In Ref.~\cite{Falkowski:2000yq}, more general gaugings are considered, 
which corresponds to the $(\alp,\bt)$-gauging. 
Its off-shell formulation is 
incorporated into the off-shell formulation in Ref.~\cite{Kugo:2002js}. 
}
As shown in Ref.~\cite{Lukas:1998yy}, 
it has a {\it linearly} warped BPS background geometry given by 
\begin{eqnarray}
ds^2 &=& (A+2 \alpha y)^{1/6}dx^2-dy^2, 
\label{eq:ds2hm}
\end{eqnarray}
where $A$ is a constant. 
In this case, the gauge parameter $\alpha$ is related to a flux, 
that is, a curvature four-form integrated over a four-cycle $C_4$ 
of the Calabi-Yau manifold, 
\begin{eqnarray}
\alpha &\propto& \int_{C_4}\, {\rm tr}\, R \wedge R. 
\end{eqnarray}

If we turn on the $\bt$-gauging, the background 
metric~(\ref{eq:ds2hm}) becomes 
\be
 ds^2 = e^{-2\bt y}(A+2\alp y)^{1/6}dx^2-dy^2, 
\ee
and the model has an {\it exponentially} warped geometry. 
Especially it is reduced to the SUSY RS 
model~\cite{Altendorfer:2000rr,Gherghetta:2000qt,Falkowski:2000yq} 
when $\alp=0$. 
This means that the $\bt$-gauging induces 
a negative cosmological constant in the bulk, 
which leads to the AdS curvature~$k=\bt$. 
We can see from the 5D action that 
it also induces the bulk mass~$m=3\bt/2$ for the physical hypermultiplet. 
By introducing the $\gm$-gauging to this case, the parameters~$k$ and $m$ 
become independent and are given by 
\begin{eqnarray}
k &=& \bt-\frac{\gm}{3}, \qquad 
m \ = \ \frac{3}{2}(\beta+\gamma). 
\label{eq:km}
\end{eqnarray}
Note that these relations hold only when $\alp=0$. 
If we turn on the $\alp$-gauging, the model deviates from the SUSY RS model, 
and the above relations are modified to more complicated ones. 

Thus the $(\alpha,\beta,\gamma)$-gauging model, which we refer to as 
the hybrid model in this paper, can be regarded 
as a hybrid formulation of the SUSY RS model and the 5D heterotic 
M-theory. 
Although we have gauged only three directions among eight isometries 
within $SU(2,1)$, 
this $(\alp,\bt,\gm)$-gauging model contains most of important structures 
in the gauged supergravity with a universal hypermultiplet
because both the linear and the exponential warp factors 
can be realized just by taking different limit of the 
the gauge parameters. 
The other gauge parameters in Eq.~(\ref{eq:u21gg}) generate a similar 
warping such as $\sinh(y)$ or $\cosh(y)$ which are just 
combinations of the exponential factors. 

Therefore, in the following, we study the 
$(\alpha,\beta,\gamma)$-gauging model in detail with boundary 
induced superpotentials. 
\ignore{
The off-shell dimensional reduction 
allows us to derive systematically a 4D effective theory of 
5D supergravity with arbitrary superpotential terms at the 
orbifold fixed point, and we adopt this reduction procedure. }
Since the boundary Lagrangian~$\cL^\vtht$ must be invariant 
for $U(1)_T$, it is written as 
\be
 \cL^\vtht = e^{3\sgm}
 \sbk{\int d^2\tht\;\Phi^2\Phi^3P_\vtht(Q)+\hc}, 
 \label{eq:lbrane}
\ee
where $P_\vtht(Q)$ ($\vtht=0,\pi$) are the boundary superpotentials. 
The induced chiral multiplet~$Q$ has zero Weyl weight and is neutral 
for $U(1)_T$. 
Thus it is identified as 
\be
 Q=\frac{\Phi^5}{\Phi^3}. 
\ee
Note that only $Z_2$-even multiplets can appear in $\cL^\vtht$.

\subsection{4D effective action}
Now we derive the 4D effective theory 
for the $(\alpha,\beta,\gamma)$-gauging 
with superpotentials $P_\vartheta$ at the orbifold fixed points. 
For this purpose, we adopt the off-shell dimensional 
reduction proposed by Refs.~\cite{Correia:2006pj,Abe:2006eg}, 
which are based on the $N=1$ superspace description~\cite{
Paccetti Correia:2004ri} of the 5D off-shell supergravity
and developed in subsequent studies~\cite{Abe:2005ac}. 
This method enables us to derive the 4D off-shell effective action directly 
from the 5D off-shell supergravity action 
{\it keeping the $N=1$ off-shell structure}. 
The procedure is straightforward. 
We start from the $N=1$ off-shell description 
of 5D action~(\ref{eq:lbulk}) with (\ref{eq:lbrane}). 
After some gauge transformation, we drop kinetic terms 
for $Z_2$-odd multiplets because they are negligible in low energies. 
Then those multiplets play a role of the Lagrange multipliers 
and their equations of motion extract zero-modes 
from the $Z_2$-even multiplets. 
Note that only the $Z_2$-even multiplets have zero-modes that appear 
in the effective theory. 

In our model the $Z_2$-even multiplets are $\Sgm^0$, $\Phi^2$, $\Phi^3$ 
and $\Phi^5$. 
Due to the $U(1)_T$-invariance, they appear in the action 
only through the combinations of $\Sgm^0$, $\Phi^2\Phi^3$ and $\Phi^5/\Phi^3$ 
which carry the zero-modes, the radion multiplet~$T$, 
4D chiral compensator~$\phi$ and the matter multiplet~$H$, respectively. 
Following the procedure of Ref.~\cite{Correia:2006pj,Abe:2006eg}, 
we obtain the 4D effective action as 
\begin{eqnarray}
S_{\rm eff} &=& -3 \int d^4\tht \, 
|\phi|^2 e^{-K/3} 
+\left\{ \int d^2 \theta \, \phi^3 W 
+\textrm{h.c.} \right\}, 
\end{eqnarray}
where the K\"{a}hler potential~$K$ and the superpotential~$W$ are 
given by 
\begin{eqnarray}
K &=& -3 \ln \int^{\pi {\rm Re}\,T}_0 dt\,
e^{-2 \beta t} \bigg\{ 
\cosh \left( 2t \sqrt{\gamma^2 -2\alpha \gamma} 
\right)\,(1-|H|^2) 
\nonumber \\ && \qquad 
+\sinh \left( 2t \sqrt{\gamma^2 -2\alpha \gamma} 
\right)\,\frac{(\alpha+\gamma)(1+|H|^2)+\alpha 
(H+\bar{H})}{\sqrt{\gamma^2 -2\alpha \gamma}}
\bigg\}^{\frac{1}{3}}, 
\label{eq:abckp} \\
W &=& \frac{1}{4}\sum_{\vartheta=0,\pi} e^{-3\vartheta \beta T} 
\left\{ c_\vartheta+(\alpha+\gamma) s_\vartheta 
+\alpha s_\vartheta H \right\} 
P_\vartheta(H_\vartheta). 
\label{eq:abcsp}
\end{eqnarray}
Here, $H_\vtht$ and $(c_\vtht,s_\vtht)$ are defined as  
\begin{eqnarray}
H_\vartheta &\equiv& 
\frac{-\alpha s_\vartheta
+\left( c_\vartheta 
-(\alpha+\gamma) s_\vartheta \right) H}{
c_\vartheta +(\alpha+\gamma) s_\vartheta 
+\alpha s_\vartheta H}, 
\end{eqnarray}
\begin{eqnarray}
c_\vartheta &\equiv& 
\cosh \left( \vartheta T \sqrt{\gamma^2 
-2\alpha \gamma} \right), \qquad 
s_\vartheta \ \equiv \ 
\frac{\sinh \left( \vartheta T \sqrt{\gamma^2 
-2\alpha \gamma} \right)}{
\sqrt{\gamma^2 -2\alpha \gamma}}. 
\end{eqnarray}

\ignore{
As we will see in the following sections, the above effective 
action reproduces the 4D effective action of heterotic 
M-theory~\cite{Lukas:1997fg} and SUSY RS model~\cite{
Gherghetta:2000qt} in the corresponding limit. }

As the boundary superpotentials~$P_\vtht(Q)$ in (\ref{eq:lbrane}), 
we consider the following polynomials. 
\be
 P_\vtht(Q) = \sum_{n \geq 0} w_\vtht^{(n)}Q^n, 
 \label{expr:P}
\ee
where $w_\vtht^{(n)}$ ($n=0,1,2,\ldots$) are constants. 

There are two simple cases to analyse. 
When $\gm=0$, the K\"{a}hler potential~$K$ 
can be expressed by an analytic function. 
When $\alp=0$, on the other hand, 
the superpotential~$W$ is reduced to 
a simple form, \ie, a polynomial for $H$. 
Thus we will discuss these two cases in detail 
in the next two sections.

\section{Moduli stabilization in hybrid model ($\bdm{\gamma = 0}$)}
\label{sec:hybrid}
In the case that $\gamma = 0$, the $t$-integration in the 
K\"ahler potential (\ref{eq:abckp}) can be carried 
out analytically. 
Such analytic expression allows us to study more about 
the 4D effective theory of the above hybrid model. 
In this section, we analyze the vacuum structure of 
the $(\alpha,\beta)$-gauging model assuming, for simplicity 
and concreteness, the boundary superpotentials consist of 
only constant and tadpole terms for the universal hypermultiplet, 
\ie, $w_\vtht^{(n)}=0$ for $n\geq 2$ in (\ref{expr:P}). 

After a K\"ahler transformation (which is equivalent to a 
rescaling of the chiral compensator~$\phi$),  
the K\"ahler potential (\ref{eq:abckp}) and the superpotential 
(\ref{eq:abcsp}) are expressed as 
\bea
 K \eql -3\ln\brc{\frac{1}{2\alp}\cF(S_0,S_\pi)}, \nonumber\\
 W \eql e^{-\frac{3}{2}qS_0}(a_0+b_0 S_0)
 -e^{-\frac{3}{2}qS_\pi}(a_\pi+b_\pi S_\pi).  \label{KW}
\eea
where $q=\beta/\alpha$ and 
\be
 \cF(S_0,S_\pi) \equiv q^{-4/3}\brc{\Gm\brkt{\frac{4}{3},q\Re S_0}
 -\Gm\brkt{\frac{4}{3},q\Re S_\pi}}. 
\ee
Here, $\Gm(a,x)=\int_x^{\infty}dt\,t^{a-1}e^{-t}$ 
is the incomplete gamma function, and the chiral multiplets 
$S_\vartheta$ are defined as 
\be
S_0 \equiv \ \frac{1-H}{1+H}, \qquad 
S_\pi \equiv S_0+2\pi\alp T.  \label{def_Svtht}
\ee
The parameters $a_\vth$ and $b_\vth$ in the superpotential 
are given by linear combinations of the constants in the boundary 
superpotentials (\ref{expr:P}) as 
\bea
 a_\vth \defa \frac{1}{8}\brkt{w_\vth^{(0)}+w_\vth^{(1)}}, \;\;\;\;\;
 b_\vth \equiv \frac{1}{8}\brkt{w_\vth^{(0)}-w_\vth^{(1)}}, 
 \;\;\;\;\; (\vth=0,\pi). 
\eea
The scalar component of $S_0$ defined in (\ref{def_Svtht}) 
corresponds to a zero-mode for the $Z_2$-even 5D scalar~$S$ 
in the notation of Refs.~\cite{attractor,Falkowski:2000yq}. 

For given $K$ and $W$, 
the scalar potential~$V$ is calculated by the formula, 
\be
 V = e^K\brc{K^{X\bar{Y}}D_X W D_{\bar{Y}}\bar{W}-3\abs{W}^2}, 
 \label{formula_for_V}
\ee
where $D_X W\equiv W_X+K_X W$, and $X,Y$ run over all chiral multiplets. 
Note that the scalar potential~(\ref{formula_for_V}) is that 
in the Einstein frame, which corresponds to a gauge where 
the chiral compensator scalar is fixed as 
\be
 \phi=e^{K/6}. \label{Einstein_frame}
\ee
Here and henceforth, we take the unit of the 4D Planck mass, 
\ie, $M_{\rm Pl}=1$. 
(In the previous section, we took the unit of the 5D Planck mass.)

In our model, $V$ is calculated as 
\bea
 V \eql \frac{-8\alp^3}{\cF(S_0,S_\pi)^2}
 \left[\brc{\frac{\cF(S_0,S_\pi)}{3}+
 \frac{(\Re S_0)^{4/3}e^{-q\Re S_0}}{1-3q\Re S_0}
 -\frac{(\Re S_\pi)^{4/3}e^{-q\Re S_\pi}}{1-3q\Re S_\pi}}^{-1} \right. 
 \nonumber\\ &&\hspace{25mm} \times\abs{
 e^{-\frac{3}{2}qS_0}\frac{a_0-b_0\bar{S}_0}{1-3q\Re S_0}
 -e^{-\frac{3}{2}qS_\pi}\frac{a_\pi-b_\pi\bar{S}_\pi}{1-3q\Re S_\pi}}^2
 \nonumber\\
 &&\hspace{25mm} \left.
-4\left\{(\Re S_0)^{2/3}e^{-2q\Re S_0}
 \frac{\abs{b_0-\frac{3}{2}q(a_0+b_0 S_0)}^2}{1-3q\Re S_0} \right.\right.
 \nonumber\\ &&\left.\left.\hspace{35mm}
 -(\Re S_\pi)^{2/3}e^{-2q\Re S_\pi}\frac{\abs{b_\pi-\frac{3}{2}q(a_\pi+b_\pi S_\pi)}^2}
 {1-3q\Re S_\pi}\right\}\right]. 
\label{eq:abv}
\eea
In this paper we use the same symbols for the scalar fields 
as the chiral multiplets they belong to.

\subsection{Heterotic M-theory limit (pure $\bdm{\alp}$-gauging)}
In the limit $\beta \to 0$, Eq.(\ref{KW}) becomes 
\bea
 K \eql -3\ln\sbk{\frac{3}{8\alp}\brc{(\Re S_\pi)^{4/3}-(\Re S_0)^{4/3}}}, 
 \nonumber\\
 W \eql b_0\brc{C+S_0-r S_\pi}, \label{KW_nbt}
\eea
where
\bea
 C \equiv \frac{a_0-a_\pi}{b_0}, \qquad 
 r \equiv \frac{b_\pi}{b_0}. 
\label{eq:stcr}
\eea

The above K\"ahler potential reproduces the known result, i.e., the 4D 
effective K\"ahler potential of the heterotic M-theory~\cite{Lukas:1997fg} 
when $\Re S_0 \gg\pi\alp\Re T$ as pointed out in Ref.~\cite{Correia:2006pj}. 
The superpotential~$W$ originates 
from the boundary superpotentials~(\ref{expr:P}). 
In this case, the vacuum values of 
$S_0$, $S_\pi$ and $\Re T$ determine 
the Calabi-Yau volume at $y=0,\pi R$ and 
the radius of the compact 11th dimension, respectively. 
Here we assume\footnote{In the case of $\alp<0$, we can 
repeat the same arguments by exchanging $S_0$ and $S_\pi$.} 
that $\alp>0$, then the scalar fields have to satisfy 
$\Re S_\pi>\Re S_0>0$ in order for them to have 
the physical interpretation as the volumes. 

The scalar potential (\ref{eq:abv}) is now reduced to 
\be
 V = \brkt{\frac{8\alp}{3}}^3\abs{b_0}^2\brc{
 \frac{\abs{C-\bar{S}_0+r\bar{S}_\pi}^2}
 {\brc{(\Re S_\pi)^{4/3}-(\Re S_0)^{4/3}}^3}
 -3\frac{\abs{r}^2(\Re S_\pi)^{2/3}-(\Re S_0)^{2/3}}{\brc{
 (\Re S_\pi)^{4/3}-(\Re S_0)^{4/3}}^2}}. 
 \label{V_nbt}
\ee

From the SUSY preserving conditions: 
$D_{S_0}W=D_{S_\pi}W=0$, 
we obtain 
\bea
 \Re S_\pi \eql r^3\Re S_0.  \nonumber\\
 2(\Re S_0)^{1/3}\brkt{C+S_0-rS_\pi} \eql
 (\Re S_0)^{4/3}-(\Re S_\pi)^{4/3}.  \label{RrR}
\eea
The first equation indicates that SUSY point exists only when $r^3$ 
is real and positive. 
Thus we assume that $r$ is real positive in the following. 
Solving (\ref{RrR}), we find a SUSY point, 
\bea
 (\Re S_0,\Re S_\pi) \eql \brkt{\frac{2C_R}{r^4-1}, 
 \frac{2C_R r^3}{r^4-1}}, \label{SUSYpoint_nbt} \\
 C_I+\Im S_0-r\Im S_\pi \eql 0, \label{Im:nbt}
\eea
where 
\be
 C \equiv C_R+iC_I. 
\ee
Since the scalars must satisfy $\Re S_\pi> \Re S_0>0$, this point is 
in the physical region only when $C_R>0$ and $r>1$. 
We focus on this parameter region. 
From Eq.~(\ref{Im:nbt}), we find a flat direction 
in an imaginary direction of $S_\vtht$. 
The superpotential takes the following value at the SUSY point.  
\be
 W = -b_0C_R. 
\ee
Thus from (\ref{formula_for_V}), 
the vacuum energy is negative at this SUSY point, 
that is, the geometry is AdS${}_4$. 
By evaluating the second derivatives of the potential~(\ref{V_nbt}), 
we can see that this SUSY point is a saddle point.  
Here we should note that SUSY points are always stable 
in a sense that they satisfy the Breitenlohner-Freedman 
bound~\cite{BFbound}.\footnote{
For a compact proof, see Appendix~C in Ref.~\cite{SUSYstable}, 
for example.}
As will be done in Sect.~\ref{uplifting}, we will uplift 
the negative vacuum energy of the SUSY AdS vacuum 
by a SUSY breaking vacuum energy in the hidden sector 
in order to obtain a SUSY breaking Minkowski vacuum. 
In general a SUSY saddle point remains to be a saddle point 
after the uplifting unless the uplifting potential is steep, 
and such saddle point after the uplifting is not stable any more. 
So we would like to look for a local minimum of the potential 
which is expected to be stable after the uplifting. 

When $r\gg 1$, there is another stationary point,  
\be
 (\Re S_0,\Re S_\pi) = \brkt{\frac{16C_R}{r^4}, \frac{2C_R}{r}}
 \brc{1+\cO\brkt{\frac{1}{r}}},  \label{nonSUSYpoint_nbt}
\ee
along the direction~(\ref{Im:nbt}). 
In contrast to the previous point~(\ref{SUSYpoint_nbt}), 
this point is a local minimum 
except for the flat direction~(\ref{Im:nbt}). 
At this point, 
\bea
 D_{S_0}W \eql -b_0, \;\;\;\;\;
 D_{S_\pi}W = b_0\times \cO\brkt{\frac{1}{r}}, \;\;\;\;\;
 W = -b_0C_R. 
\eea
Thus this is a SUSY breaking AdS${}_4$ vacuum. 

\ignore{
Finally we check the stability of the SUSY preserving 
and breaking stationary points, (\ref{SUSYpoint_nbt}) 
and (\ref{nonSUSYpoint_nbt}). 
The second derivatives of (\ref{eq:tlv}) are obtained as 
(\ref{eq:d2tlv}). At the SUSY point~(\ref{SUSYpoint_nbt}), 
we find the Hessian matrix, 
\be
 X \equiv \begin{pmatrix} \frac{\der^2\tl{V}}{\der s_0^2} & 
 \frac{\der^2\tl{V}}{\der s_0\der s_\pi} \\
 \frac{\der^2\tl{V}}{\der s_\pi\der s_0} & \frac{\der^2\tl{V}}{\der s_\pi^2}
 \end{pmatrix} 
 = \frac{r^4-1}{48C_R^4}\begin{pmatrix} 15-r^4 & -14r \\ -14r & 15r^2-r^{-2}
 \end{pmatrix}, 
\ee
and 
\be
 \det X = -15\brkt{\frac{r^4-1}{48C_R^4}}^2\brkt{\frac{r^4-1}{r}}^2 <0. 
\ee
Thus the SUSY point turns out to be a saddle point. 
On the other hand, at the SUSY breaking stationary point 
(\ref{nonSUSYpoint_nbt}), we find 
\be
 X \simeq \frac{r^5}{C_R^4}
 \begin{pmatrix} \frac{r^3}{384} & -\frac{17}{24} \\
 -\frac{17}{24} & \frac{5r}{16} \end{pmatrix}, 
\ee
and the SUSY breaking stationary point is a local minimum 
when $r\gg 1$. 
}

\subsection{Randall-Sundrum limit (pure $\bdm{\bt}$-gauging)}
In the limit $\alpha \to 0$, the K\"ahler and the 
superpotentials (\ref{KW}) become 
\bea
 K \eql -3\ln\brkt{\frac{1-\abs{\Omg}^2}{2\adsc}}
 -\ln\brkt{\Re S_0}, \nonumber\\
 W \eql (a_0+b_0 S_0)-(a_\pi+b_\pi S_0)\Omg^3, 
 \label{KW_nalp}
\eea
where $\Omg \equiv e^{-\adsc\pi T}$ is a warp factor superfield. 
The above $K$ reproduces the radion K\"ahler potential 
of the SUSY RS model~\cite{Luty:2000ec}. 
In the following, we assume $\adsc>0$. 
Then the scalar fields must satisfy $\Re S_0>0$ and $\abs{\Omg}<1$. 
Although $H$ is more conventional than $S_0$ for the SUSY RS model, 
we use $S_0$ as a matter chiral multiplet because we will 
interpolate this model and the Heterotic M-theory limit 
($\alp\neq 0$, $\bt\to 0$). 
We can always translate $S_0$ to $H$ by the relation~(\ref{def_Svtht}). 

In this limit, the scalar potential (\ref{eq:abv}) is 
reduced to 
\bea
 V \eql \frac{8\adsc^3}{(1-\abs{\Omg}^2)^2\Re S_0}
 \bigg\{ 
 \frac{\abs{(a_0-a_\pi\Omg^3)-\bar{S}_0(b_0-b_\pi\Omg^3)}^2}{1-\abs{\Omg}^2}
 \nonumber \\ && \qquad\qquad\qquad\qquad 
 -3\brkt{\abs{a_0+b_0 S_0}^2-\abs{\Omg}^4\abs{a_\pi+b_\pi S_0}^2}
 \bigg\}. 
\label{eq:bv}
\eea
\ignore{
From (\ref{KW_nalp}), we obtain 
the K\"ahler covariant derivatives of the superpotential 
$D_IW=W_I+K_IW$ as 
\bea
 D_S W \eql \frac{-(a_0-a_\pi\Omg^3)+\bar{S}(b_0-b_\pi\Omg^3)}
 {S+\bar{S}}, \nonumber\\
 D_\Omg W \eql \frac{3\brc{\bar{\Omg}(a_0+b_0 S)-\Omg^2(a_\pi+b_\pi S)}}
 {1-\abs{\Omg}^2}. 
\eea}
From the SUSY conditions:~$D_{S_0} W=D_\Omg W=0$, we obtain  
\be
 \brkt{a_0-a_\pi\Omg^3}\brkt{\bar{b}_0-\bar{b}_\pi\frac{\bar{\Omg}^2}{\Omg}}
 +\brkt{b_0-b_\pi\Omg^3}\brkt{\bar{a}_0-\bar{a}_\pi\frac{\bar{\Omg}^2}{\Omg}} = 0. 
 \label{SUSYcond_omg}
\ee
In the case that $a_0b_\pi-b_0a_\pi\neq 0$, we find a SUSY solution as 
\be
 S_0=\frac{\bar{a}_0-\bar{a}_\pi\bar{\Omg}^3}{\bar{b}_0-\bar{b}_\pi\bar{\Omg}^3}
 = -\frac{a_0-a_\pi(\Omg^2/\bar{\Omg})}{b_0-b_\pi(\Omg^2/\bar{\Omg})}, 
\ee
where $\Omg$ is a solution of Eq.~(\ref{SUSYcond_omg}). 

On the other hand, in the case that $a_0b_\pi-b_0a_\pi=0$, that is, 
\begin{eqnarray}
(a_0,a_\pi) &=& c\,(b_0,b_\pi), \;\;\;\;\;
 \brkt{\mbox{or }
 (w_0^{(0)},w_\pi^{(0)}) = \frac{c+1}{c-1}(w_0^{(1)},w_\pi^{(1)})}
\label{eq:bgsd}
\end{eqnarray}
where $c$ is a constant, the relation 
(\ref{SUSYcond_omg}) is rewritten as 
\be
 (\Re c)(1-r\Omg^3)\brkt{1-\bar{r}\frac{\bar{\Omg}^2}{\Omg}} = 0, 
\ee
where $r\equiv b_\pi/b_0$, and the superpotential in this case 
is found as 
\be
 W=b_0(c+S_0)(1-r\Omg^3). 
\ee

%
For $\Re c<0$, the SUSY point is found as 
\be
 (S_0,\Omg) = \brkt{-c,\,r^{-1/3}}.  
 \label{SUSYpoint1_nalp}
\ee
At this point, $W=0$ and thus the vacuum energy vanishes, 
resulting a local Minkowski minimum. 
This corresponds to the SUSY Minkowski vacuum discussed 
in Ref.~\cite{Maru:2003mq}, in which 
the boundary superpotentials~(\ref{expr:P}) consist of 
only the tadpole terms, \ie, $w_\vtht^{(0)}=0$ (or $c=-1$.)

For $\Re c>0$, the SUSY solution is found as 
\be
 (S_0,\Omg) = \brkt{\bar{c},\, \abs{r}^{-4/3} \bar{r}^{1/3}}, 
 \label{SUSYpoint2_nalp}
\ee
where the superpotential does not vanish at this point, 
\be
 W = 2b_0(\Re c)\brkt{1-\frac{1}{\abs{r}^2}} \neq 0. 
\ee
This means that the vacuum energy is negative and 
the geometry becomes AdS${}_4$. 
This SUSY solution is a saddle point.\footnote{
A case that only constant terms exist 
in the boundary superpotentials, \ie, $c=1$, is studied 
in Ref.~\cite{Maru:2006ji} where a consistent result 
with ours is obtained. 
}

In either case, the SUSY point exists in the region 
$\Re S_0>0$ and $\abs{\Omg}<1$ only when $\abs{r}>1$. 
For $\Re c=0$, $S=-c=\bar{c}$ and $\Omg$ is undetermined. 

In the following, we focus on the case (\ref{eq:bgsd}) 
where the scalar potential (\ref{eq:bv}) is simplified as  
\be
 V = \frac{8\adsc^3\abs{b_0}^2}{(1-\abs{\Omg}^2)^2\Re S_0}
 \brc{\frac{\abs{c-\bar{S}_0}^2\abs{1-r\Omg^3}^2}{1-\abs{\Omg}^2}
 -3\abs{c+S_0}^2\brkt{1-\abs{r}^2\abs{\Omg}^4}}.  \label{V_Somg}
\ee
Here we have decomposed the complex scalars and the parameters 
into real ones as 
\be
 S_0 = s+i\sgm, \;\;\;\;\;
 \Omg = \omg e^{i\vph}, \;\;\;\;\;
 r = \rho e^{i\zeta}, \;\;\;\;\;
 c = c_R+ic_I. 
\ee
From the stationary conditions for $\sgm$ and $\vph$, we obtain 
\be
 c_I+\sgm=0, \;\;\;\;\;
 \sin(\zeta+3\vph)=0. 
\ee
\ignore{
In fact the SUSY points~(\ref{SUSYpoint1_nalp}) 
and (\ref{SUSYpoint2_nalp}) satisfy these conditions. 
When $r\gg 1$, there are other stationary points than these SUSY points, 
\be
 (s,\omg) \simeq \brkt{\frac{1 \mp 2^{5/8}3^{-1/8}\rho^{1/4}}
 {1 \pm 2^{5/8}3^{-1/8}\rho^{1/4}}c_R,\,
 \brkt{\frac{2}{3}}^{1/4}\rho^{-1/2}}. 
\ee
These are saddle points. 
After fixing $\sgm$ and $\vph$ by these conditions, 
the derivatives of the scalar potential are obtained as 
\bea
 \frac{\der\tl{V}}{\der s} \eql \frac{c_R^2-s^2}{(1-\omg^2)^3s^2}
 \brc{2-3\omg^2+2\rho\omg^3-3\rho^2\omg^4+2\rho^2\omg^6}, \nonumber\\
 \frac{\der\tl{V}}{\der\omg} \eql \frac{6\omg(1-\rho\omg)}{(1-\omg^2)^4s}
 \brc{(c_R-s)^2(1-\rho\omg^3)-2(c_R+s)^2(1+\rho\omg)(1-\omg^2)}, 
\eea
where $\tl{V}=V/(8\adsc^3\abs{b_0}^2)$. 
We easily find that the SUSY points~(\ref{SUSYpoint1_nalp}) 
and (\ref{SUSYpoint2_nalp}) satisfy the stationary condition 
$\der V/\der s=\der V/\der\omg=0$. In addition, 
there are another stationary solutions, 
\be
 (s,\omg) \simeq \brkt{\frac{1 \mp 2^{5/8}3^{-1/8}\rho^{1/4}}
 {1 \pm 2^{5/8}3^{-1/8}\rho^{1/4}}c_R,\,
 \brkt{\frac{2}{3}}^{1/4}\rho^{-1/2}}. 
\ee
Next we study the stability of the SUSY stationary solutions. 
The second derivatives of $\tl{V}=V/(8\adsc^3\abs{b_0}^2)$ 
are found as (\ref{eq:d2bv}). At the SUSY point 
(\ref{SUSYpoint1_nalp}) where $(s,\omg)=(-c_R,\rho^{-1/3})$, 
these derivatives are given by 
\bea
 \frac{\der^2\tl{V}}{\der s^2} \eql -\frac{6\rho^{4/3}(1+\rho^{-2/3})^2}
 {(1-\rho^{-2/3})c_R}, \quad 
 \frac{\der^2\tl{V}}{\der s\der\omg} = 0, \quad 
 \frac{\der^2\tl{V}}{\der\omg^2} = -\frac{72c_R\rho^{4/3}(1+\rho^{-2/3})}
 {(1-\rho^{-2/3})^3}.  
 \label{der2V_SUSY1}
\eea
Therefore, this point is a local minimum for $c_R<0$ and $\rho>1$. 
On the other hand, at the other SUSY point~(\ref{SUSYpoint2_nalp}) 
where $(s,\omg)=(c_R,\rho^{-1})$, we find 
\bea
 \frac{\der^2\tl{V}}{\der s^2} \eql -\frac{4}{(1-\rho^{-2})c_R}, 
\quad 
 \frac{\der^2\tl{V}}{\der s\der\omg} = 0, \quad 
 \frac{\der^2\tl{V}}{\der\omg^2} = \frac{96c_R}{(1-\rho^{-2})^3}. 
\eea
Therefore, this is a saddle point. 
}

Focusing on the SUSY local minimum (\ref{SUSYpoint1_nalp}), 
we calculate the mass eigenvalues. 
Evaluating the second derivatives of the scalar potential~(\ref{V_Somg}), 
we can see that the four real scalars~$(s,\omg,\sgm,\vph)$ do not mix 
with each other. 
Then after normalizing them canonically, the mass eigenvalues are 
found as 
\bea
 m_s^2 \eql m_\omg^2 = 96\adsc^3\abs{b_0}^2\abs{c_R}
 \frac{\rho^{4/3}(1+\rho^{-2/3})^2}{1-\rho^{-2/3}}, \nonumber\\
 m_\sgm^2 \eql m_\vph^2 = 48\adsc^3\abs{b_0}^2\abs{c_R}
 \frac{\rho^{2/3}}{1-\rho^{-2/3}}. 
 \label{moduli_masses}
\eea
We have assumed $c_R<0$ and $\rho>1$.

\subsection{Interpolation}
In the previous subsections, we analyzed in detail the vacuum 
structures in the two typical limits of the hybrid model, 
\ie, the heterotic M-theory limit (pure $\alp$-gauging) and 
the SUSY RS limit (pure $\bt$-gauging). 
In this subsection, we study intermediate regions 
of the hybrid model. 
We assume that the parameters satisfy 
the relation~(\ref{eq:bgsd}) with $c_R\equiv\Re c<0$ 
for simplicity of the analysis.

\subsubsection{Numerical result} \label{numerical}
\ignore{
\begin{figure}[t]
\begin{center}
\epsfig{file=st0.eps,width=0.4\linewidth}
\end{center}
\caption{The position of the SUSY point 
in the $(\Re S,\Re T)$-plane. 
We vary the parameter $\alpha$ from $0.1$ (north-east) 
to $100$ (south-west) by fixing $\beta=1$, $c=-40$, 
$r=10^5$ and $b_0=10^{-14}$. All the mass scales are 
measured in the unit $M_{Pl}=1$.}
\label{fig:st0}
\end{figure}
%
The position of the SUSY point is plotted in the 
$(S,T)$-plane in Fig.~\ref{fig:st0} by varying $\alpha$ from $0.1$ 
(RS limit) to $100$ (M-theory limit) with the other parameters fixed, 
e.g., $\beta=1$. Since the vacuum value of $T$ represents 
the length which should be larger than the Planck length, the 
region close to the RS limit is favored in Fig.~\ref{fig:st0}. 
This is indeed a generic fact which can be found from the 
expressions (\ref{SUSYpoint1_nalp}) and (\ref{SUSYpoint_nbt}). 
}

First we show some numerical results. 
We focus on the SUSY preserving points~(\ref{SUSYpoint_nbt}) 
in the pure $\alp$-gauging and 
(\ref{SUSYpoint1_nalp}) in the pure $\bt$-gauging, 
and numerically interpolate these stationary points 
in the intermediate region, 
where both $\alpha$ and $\beta$ are nonvanishing. 
\begin{figure}[t]
\begin{center}
\ \\ \vspace{-1cm}
\epsfig{file=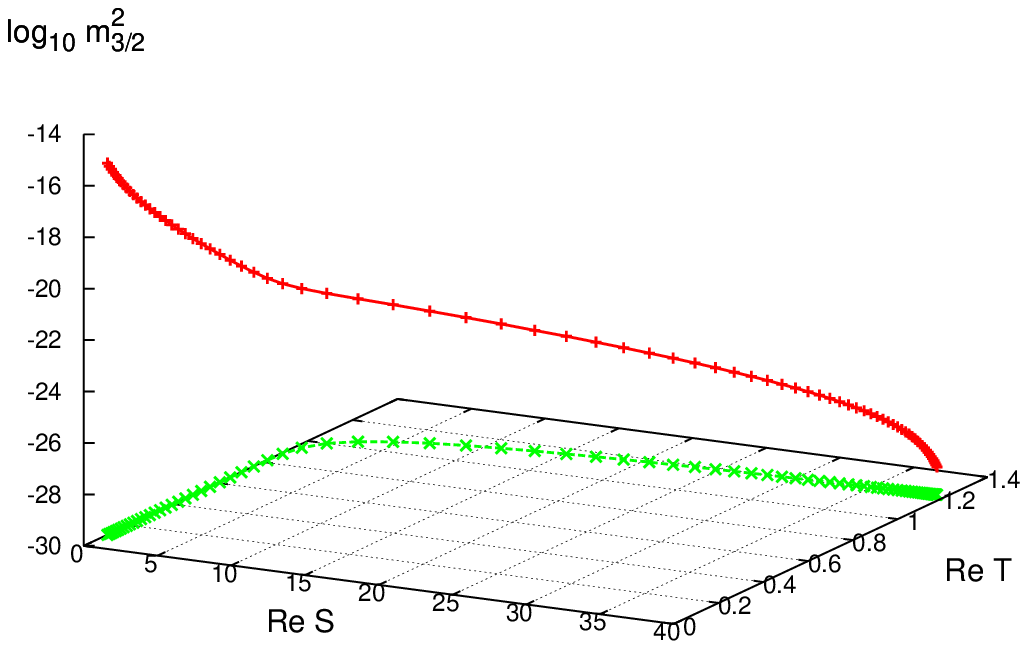,width=0.9\linewidth} \\
\vspace{-1.5cm}
\end{center}
\caption{The gravitino mass on the $(\Re S,\Re T)$-plane. 
We vary the parameter $\alpha$ from $100$ (left end-point) 
to $0.1$ (right end-point) by fixing $\beta=1$, $c=-40$, 
$r=10^5$ and $b_0=10^{-14}$. All the mass scales are 
measured in the unit $M_{\rm Pl}=1$.
}
\label{fig:stm}
\end{figure}
Fig.~\ref{fig:stm} plots the gravitino mass and 
the position of the SUSY point on the $(\Re S,\Re T)$ plane 
for various values of $\alp$ with the other parameters fixed. 
The left end-point of the curve corresponds to $\alp=100$ 
(near the M-theory limit) and the other end-point 
corresponds to $\alp=0.1$ (near the SUSY RS limit). 
The curve on the $(\Re S,\Re T)$ plane represents the projection 
on that plane. 
Since the vacuum energy at the SUSY point~$V_{\rm SUSY}$ is 
related to the gravitino mass~$m_{3/2}=e^{K/2}\abs{W}$ through 
$V_{\rm SUSY}=-3m_{3/2}^2$, we can see from Fig.~\ref{fig:stm} 
that $\abs{V_{\rm SUSY}}$ monotonically decreases 
as $\alpha$ decreases. 
This is consistent with the fact that 
the SUSY point~(\ref{SUSYpoint1_nalp}) is a 
Minkowski vacuum in the RS limit. 
Since $\Re T$ corresponds to the size of the orbifold, 
it should be larger than the Planck length~$M_{\rm Pl}^{-1}=1$. 
Thus we can see from Fig.~\ref{fig:stm} 
that the region around the RS limit is favored 
for our parameter choice. 

\begin{figure}[t]
\begin{center}
\ \\ \vspace{-1.5cm}
\epsfig{file=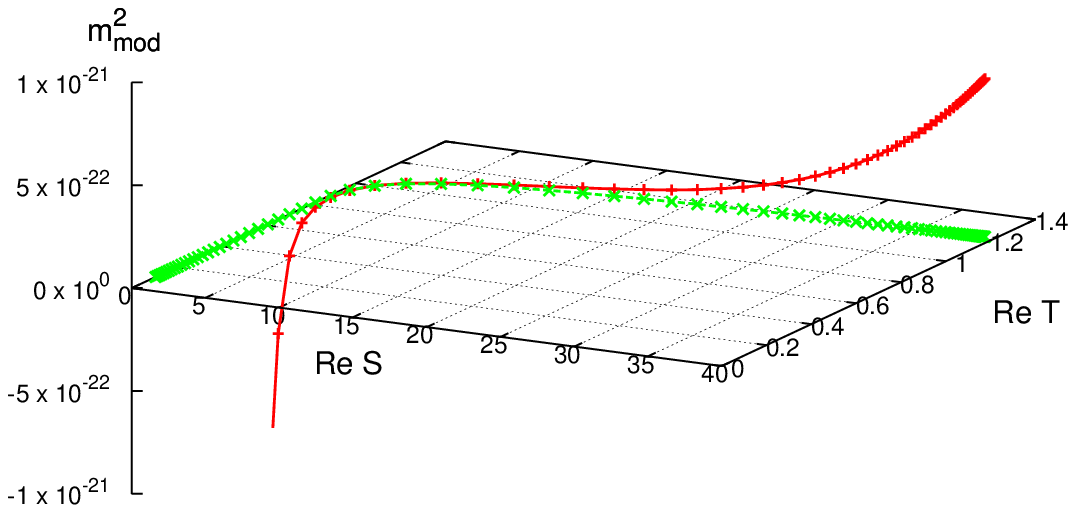,width=0.9\linewidth} \\
\vspace{-2.5cm}
\end{center}
\caption{The lightest modulus mass squared 
on the $(\Re S,\Re T)$-plane. 
The parameters are the same as in Fig.~\ref{fig:stm}. 
The mass squared~$m_{\rm mod}^2$ turns to be positive at $\alp\simeq 7$. 
}
\label{fig:stm2}
\end{figure}
Fig.~\ref{fig:stm2} plots the lightest modulus mass 
squared~$m_{\rm mod}^2$ 
and the SUSY point on the $(\Re S,\Re T)$ plane. 
Again we vary $\alp$ from 100 (left end-point) to 0.1 (right end-point) 
while fix the other parameters. 
Now we can see from this figure that the SUSY point near the 
M-theory limit (left end-point) is not local minimum 
since the lightest modulus is tachyonic, but this tachyonic mass squared 
monotonically increases as $\alp$ decreases and becomes positive 
when $\abs{c_R}\bt/\alp\gg 1$. 
Thus the region around the SUSY RS limit 
($\abs{c_R}\bt/\alp \gg 1$), where $\Re T>1$ is realized 
without any tachyonic masses, 
is the best candidate for the KKLT-type uplifting. 
We will study this region analytically in the following.

\subsubsection{Analytic result near the RS limit} \label{large_q}
In the vicinity of $\alp=0$, \ie, the SUSY RS model, 
the K\"ahler and the superpotentials in (\ref{KW}) 
are expressed as 
\bea
 K 
 \eql -3\ln\sbk{\frac{1}{2\bt}\brc{1-\abs{\Omg}^2
 +\frac{1-\abs{\Omg}^2+\abs{\Omg}^2\ln\abs{\Omg}^2}{3q\Re S_0}
 +\cO\brkt{\frac{1}{q^2(\Re S_0)^2}}}}  -\ln(\Re S_0),  \nonumber\\
 W \eql b_0\brc{(c+S_0)(1-r\Omg^3)+\frac{2r}{q}\Omg^3\ln\Omg}. 
\eea
Here $q\equiv\bt/\alp$ times $\abs{c_R}=\abs{\Re c}$ is supposed to be large. 
Then from the SUSY conditions:~$D_{S_0}W=D_\Omg W=0$,  
we find the SUSY point as 
\bea
 S_0 \eql -c\brc{1-\frac{2}{3qc} 
\brkt{1-\frac{\ln r}{1-\rho^{-2/3}}}+\cO\brkt{(qc_R)^{-2}}}, 
 \nonumber\\
 \Omg \eql r^{-1/3}\brc{1-\frac{\ln r}{9qc_R}+\cO\brkt{(qc_R)^{-2}}}. 
 \label{SUSYpoint_ab}
\eea
Here $r\equiv b_\pi/b_0$ and $\rho\equiv \abs{r}$. 
Since the SUSY point in the pure $\bt$-gauging~(\ref{SUSYpoint1_nalp}) 
is a local minimum of the potential, this point is also 
a local minimum when $q\abs{c_R}\gg 1$. 
Due to the correction from the pure $\bt$-gauging case, 
the superpotential~$W$ does not vanish at this point. 
\be
 W = -\frac{2b_0\ln r}{3q}\brc{1+\cO\brkt{(qc_R)^{-2}}}. 
\ee
Thus the vacuum energy becomes negative, evaluated as 
\be
 V = -3e^{K}\abs{W}^2 
 = -\frac{32\adsc^3\abs{b_0 \ln r}^2}{
3q^2\abs{c_R}(1-\rho^{-2/3})^3}\brc{1+\cO\brkt{\frac{1}{qc_R}}}. 
 \label{minimum_V}
\ee
Namely this is an AdS${}_4$ SUSY vacuum.

\subsection{Uplifting} \label{uplifting}
So far we have found some stationary solutions of the scalar potential 
in the hybrid model assuming certain simple superpotential 
terms at the orbifold fixed points. 
In the heterotic M-theory limit (pure $\alp$-gauging), 
the SUSY point is a saddle point. 
In the SUSY RS limit (pure $\bt$-gauging), on the other hand, 
the local minimum of the potential is a SUSY Minkowski vacuum. 
 
Finding a SUSY breaking Minkowski minimum, which is a candidate of 
our present universe, is indeed a hard task in supergravity models. 
The hybrid model studied in this section is also the case. 
As mentioned in the introduction, the KKLT model provides 
an interesting and systematic way of achieving a 
SUSY breaking minimum with vanishing vacuum energy, 
that is, uplifting SUSY AdS minimum 
by a SUSY breaking vacuum energy which is 
assumed to be well sequestered from the light moduli 
as well as the visible sector. 
Here we consider the uplifting of the AdS SUSY minimum 
in our hybrid model by a SUSY 
breaking sector which is assumed to be well sequestered from 
$S_0$ and $\Omega$ (or $T$). 

As mentioned at the end of Sec.~\ref{numerical}, 
the region around the SUSY RS model ($q\abs{c_R}\gg 1$) is the best 
candidate for the uplifting. 
So we consider such a parameter region which we discussed 
in Sec.~\ref{large_q}. 
Following the KKLT model, the uplifting potential~$U$ 
is assumed as~\cite{Choi:2004sx,Choi:2007yz} 
\bea
 U \eql 
 \int\dr^4\tht(\bar{\phi}\phi)^n\kp\tht^2\bar{\tht}^2 
 \ = \ \kp e^{nK/3} 
 \nonumber\\ 
 \eql \frac{\kp(2\adsc)^n}{(\Re S_0)^{n/3}(1-\abs{\Omg}^2)^n}
 \brc{1+\cO\brkt{\frac{1}{qc_R}}},  
\eea
where $\kp$ is a constant. 
The typical value of $n$ for the 
sequestered SUSY breaking source is given 
by $n=2$~\cite{Choi:2004sx,Choi:2007yz}. 
The total scalar potential is then given by $V_{\rm tot}\equiv V+U$. 
Then the local minimum $(S,\Omg)$ is shifted from 
the SUSY point~(\ref{SUSYpoint_ab}) by 
\bea
 \dlt S_0 \eql 
 \frac{2n\abs{\ln r}^2}{27q^2(-c_R)\rho^{4/3}(1-\rho^{-4/3})^2}
 \brc{1+\cO\brkt{\frac{1}{qc_R}}}, \nonumber\\
 \dlt\Omg \eql -\frac{2n\abs{\ln r}^2}{54q^2(-c_R)^2
 \rho^{5/3}(1-\rho^{-4/3})^2}
 \brc{1+\cO\brkt{\frac{1}{qc_R}}}. 
\eea
Here we have chosen $\kp$ as 
\be
 \kp = \frac{4(2\adsc)^{3-n}\abs{b_0\ln r}^2}{3q^2(-c_R)^{1-n/3}
 (1-\rho^{-2/3})^{3-n}}\brc{1+\cO\brkt{\frac{1}{qc_R}}}, 
 \label{exp_xi}
\ee
so that $V_{\rm tot}=0$ at the leading order 
in the $(qc_R)^{-1}$-expansion. 

Now we evaluate the F-terms of the chiral multiplets by the formulae, 
\be
 F^X = -e^{K/2}K^{X\bar{Y}}D_{\bar{Y}}\bar{W}, \;\;\;\;\;
 F^\phi = \frac{1}{3}e^{K/6}K_X F^X+e^{2K/3}\bar{W}, \label{exp_Fs}
\ee
for $X,Y=S,\Omg$. 
These provides the SUSY breaking order parameters 
for the moduli mediation ($F^X$) and for the anomaly mediation ($F^\phi$). 
They are estimated at the uplifted Minkowski minimum as 
\bea
 F^S \eql \cO(q^{-3}c_R^{-2}), \nonumber\\
 F^\Omg 
 \eql -\frac{2ne^{-i\zeta/3}(2\adsc)^{3/2}\bar{b}_0\abs{\ln r}^2}
 {27q^2(-c_R)^{3/2}\rho(1+\rho^{-2/3})^2(1-\rho^{-2/3})^{5/2}}
 \brc{1+\cO\brkt{\frac{1}{qc_R}}}, 
 \nonumber\\ 
 F^\phi \eql -\frac{2\bar{b}_0(2\adsc)^2\ln\bar{r}}{3q(-c_R)^{2/3}
 (1-\rho^{-2/3})^2}\brc{1+\cO\brkt{\frac{1}{qc_R}}}. 
\eea
If we define the anomaly/modulus ratio of SUSY breaking 
as~\cite{Choi:2004sx,Choi:2007yz}  
\be
\alpha_{A/M} \equiv \frac{1}{\ln (M_{\rm Pl}/m_{3/2})} \cdot 
\frac{F^\phi/\phi}{F^T/(T+\bar{T})}, 
\label{eq:atmratio}
\ee
we find in this case as 
\be
\alpha_{A/M} = \frac{qc_R}{\ln(M_{\rm Pl}/m_{3/2})}
 \brc{\frac{6\ln\rho}{n\ln r}\rho^{2/3}(1+\rho^{-2/3})^2
(1-\rho^{-2/3})+\cO\brkt{\frac{1}{qc_R}}}. 
 \label{expr:alp}
\ee
Since $\rho\equiv\abs{r}$ is related to $\rho=e^{3\pi\bt\Re T}$ 
from (\ref{SUSYpoint1_nalp}), we can see from (\ref{expr:alp}) that 
$\abs{\alp_{A/M}}\gg 1$ unless $\bt$ is small. 
(Notice that $\Re T$ should be larger 
than the Planck length~$M_{\rm Pl}^{-1}=1$.)
Thus the anomaly mediation tends to be dominant in this model. 
However, for small values of $\bt$, the parameter~$\rho$ 
is allowed to be of $\cO(1)$ and 
the modulus mediated contribution can be comparable to 
that of the anomaly mediation. 
For example, $\abs{\alp_{A/M}}\simeq 1$ 
when $n=2$, $r=2$, $qc_R=-8$, $\ln(M_{\rm Pl}/m_{3/2})=4\pi^2$. 
In this case, the mirage mediation is realized. 
Finally note that the moduli masses, which are given by 
(\ref{moduli_masses}) at the leading of the $(qc_R)^{-1}$-expansion, 
are much larger than the gravitino mass, 
\be
 m_{3/2}^2 = e^K\abs{W}^2 = \frac{32\bt^3\abs{b_0\ln r}^2}
 {9q^2\abs{c_R}(1-\rho^{-2/3})^3}\brc{1+\cO\brkt{\frac{1}{qc_R}}}. 
\ee

\section{Generalized Luty-Sundrum model ($\bdm{\alpha = 0}$)}
\label{sec:gls}
In the previous section we have studied the hybrid model 
in the case that $\gm=0$, 
assuming simple boundary superpotentials. 
In this section we consider a case that $\alpha = 0$ 
while $\bt$ and $\gm$ are arbitrary, in which  
the hybrid model (\ref{eq:abckp}) and (\ref{eq:abcsp}) is 
reduced to the SUSY RS model with an AdS curvature $k$ and an 
{\it arbitrary} bulk mass $m$ defined in Eq.~(\ref{eq:km}) 
for the hypermultiplet. 
We now introduce {\it generic} perturbative superpotential 
terms~(\ref{expr:P}) at the orbifold fixed points. 
The K\"ahler potential and the superpotential in this case 
are given by 
\begin{eqnarray}
K &=& 
-3 \ln \int^{\pi\Re T}_0 dt\, 
e^{-2kt} 
(1-e^{(3k-2m)t}|H|^2)^{\frac{1}{3}}, \\
W &=& 
\frac{1}{4}\sum_{\vartheta=0,\pi} e^{-3\vartheta kT} 
\sum_n w_\vartheta^{(n)} e^{n\vartheta\brkt{\frac{3}{2}k-m}T} H^n. 
\end{eqnarray}
In the hypersurface $H=0$, this corresponds to 
the Luty-Sundrum model~\cite{Luty:2000ec},\footnote{
In the original Luty-Sundrum model, the potential 
minimization was performed not in the Einstein 
frame but in the conformal frame where $\phi=1$. 
(See (\ref{Einstein_frame}).)
This leads to a different result from ours 
because the potential minimum has a negative 
vacuum energy as shown below.}  
\begin{eqnarray}
K \Big|_{H=0} \eql
-3 \ln \frac{1-e^{-k(T+\bar{T})}}{2k},  \qquad 
W \Big|_{H=0} 
= \ \frac{1}{4}\brkt{w_0^{(0)}+w_{\pi}^{(0)}e^{-3kT}}. 
\label{eq:ls}
\end{eqnarray}
We would like to mention that 
if we take the limit $k \to 0$ in the K\"ahler potential 
while keeping a finite $k$ in the superpotential in (\ref{eq:ls}), 
the model becomes equivalent to the KKLT model.

\subsection{Supersymmetry condition}
In the following we assume that 
$|w_\vtht^{(2)}|$ is sufficiently large. 
Then $H$ would receive a heavy SUSY mass around $H=0$ 
and can be integrated out by $D_HW=0$, i.e., $H=0$, 
without affecting the low energy dynamics of $T$~\cite{Abe:2006xi}. 
Then the low energy effective theory is given by the 
Luty-Sundrum model (\ref{eq:ls}). 

The SUSY conditions:~$D_HW=D_TW=0$ 
in the $H=0$ slice are given by 
\begin{eqnarray}
D_H W \Big|_{H=0} &=& 
\frac{1}{4}\brkt{w_0^{(1)}+e^{-\pi\brkt{\frac{3}{2}k+m}T}w_\pi^{(1)}} 
\ = \ 0, 
\label{eq:dhw} \\
D_T W \Big|_{H=0} &=& 
\frac{-3\pi k}{4(|e^{\pi kT}|^2-1)} 
\left( w_0^{(0)} + e^{-3\pi kT}  
|e^{\pi kT}|^2 w_\pi^{(0)} \right) 
\ = \ 0. 
\label{eq:dtw}
\end{eqnarray}
In the following we further assume 
$\Im w_\vtht^{(0)}=\Im w_\vtht^{(1)}=0$ ($\vtht=0,\pi$) 
for simplicity. Then a solution for 
(\ref{eq:dhw}) and (\ref{eq:dtw}) is found as 
\begin{eqnarray}
T &=& \bar{T} \ = \ 
\ln \left( -\frac{w_0^{(1)}}{w_\pi^{(1)}} 
\right)^{-\frac{2}{\pi(3k+2m)}} 
\ = \ 
\ln \left( -\frac{w_0^{(0)}}{w_\pi^{(0)}} 
\right)^{-\frac{1}{\pi k}}. 
\label{eq:sspgen}
\end{eqnarray}
In order for the above solution to be valid, 
the following relation must be satisfied. 
\begin{eqnarray}
-\frac{w_0^{(1)}}{w_\pi^{(1)}} 
&=& 
\left( -\frac{w_0^{(0)}}{w_\pi^{(0)}} 
\right)^{\frac{3k+2m}{2k}}. 
\label{eq:pararel}
\end{eqnarray}
One of the simplest choices satisfying (\ref{eq:pararel}) is 
\begin{eqnarray}
3k+2m &=& 0, \qquad w_0^{(1)}+w_\pi^{(1)} \ = \ 0. 
\label{eq:slpararel}
\end{eqnarray}
Because the extension to the other cases is 
straightforward, we focus on the case~(\ref{eq:slpararel})  
in the following. 
Then the SUSY point is summarized as 
\begin{eqnarray}
H &=& 0, \qquad 
T \ = \ T_{\rm SUSY} \ \equiv \ 
\ln \left( -\frac{w_0^{(0)}}{w_\pi^{(0)}} 
\right)^{-\frac{1}{\pi k}} 
\label{eq:ssp}
\end{eqnarray}

\ignore{
In general, the second derivatives of the scalar potential 
at the SUSY point $G_I=0$ are given by~\cite{Abe:2005pi} 
\begin{eqnarray}
V_{I\bar{J}} &=& 
e^G (G^{K \bar{L}} G_{IK} G_{\bar{J} \bar{L}}-2 G_{I\bar{J}}), 
\\
V_{IJ} &=& -e^G G_{IJ}, 
\end{eqnarray}
where $G=K+ \ln|W|^2$, $G_I=\partial_IG=W^{-1}D_IW$ and so on. }

In our model we find from (\ref{formula_for_V}) that 
\begin{eqnarray}
V_{H\bar{H}} &=& 
e^K K_{H\bar{H}}^{-1} (|W_{HH}|^2-2K_{H\bar{H}}|W|^2), 
\\
V_{HH} &=& -e^K \bar{W} W_{HH}, 
\\
V_{H\bar{T}} &=& V_{HT} \ = \ 0, 
\end{eqnarray}
at the SUSY point (\ref{eq:ssp}). 
The mass-square eigenvalues of $({\rm Re}\,H,\,{\rm Im}\,H)$ 
are then given by 
\begin{eqnarray}
m_{H\pm}^2 &=& K_{H\bar{H}}^{-1}(V_{H\bar{H}} \pm |V_{HH}|). 
\end{eqnarray}
In order for the SUSY point (\ref{eq:ssp}) 
to be a local minimum, a condition $m^2_{H\pm} > 0$ 
has to be satisfied, which results in 
\begin{eqnarray}
\left| \frac{W_{HH}}{W} \right| &>& 
\sqrt{\frac{1}{4}K_{H\bar{H}}^2+2K_{H\bar{H}}}
+\frac{1}{2}K_{H\bar{H}}, 
\label{eq:positive}
\end{eqnarray}
where 
\begin{eqnarray}
\frac{W_{HH}}{W} &=& 
\frac{w_\pi^{(2)}+e^{-3\pi k T}w_0^{(2)}}{
w_0^{(0)}+e^{-3\pi k T} w_\pi^{(0)}} \cdot 
2e^{3\pi k T}
\\
K_{H\bar{H}} &=& 
\frac{1-e^{-4 \pi k T}}{
1-e^{-2 \pi k T}} \cdot 
\frac{1}{2} 
e^{4 \pi k T}, 
\end{eqnarray}
and $e^{\pi k T}=-w_\pi^{(0)}/w_0^{(0)}$. 

In the case~$e^{\pi k T} \gg 1$, 
the condition (\ref{eq:positive}) becomes 
\begin{eqnarray}
\left| \frac{w_\pi^{(2)}}{w_0^{(0)}} \right| &>& 
\frac{1}{4} 
e^{\pi k T}
\ = \ \frac{1}{4} \cdot 
\frac{-w_\pi^{(0)}}{w_0^{(0)}}, 
\end{eqnarray}
which can be satisfied by, e.g., 
\begin{eqnarray}
|w_\pi^{(2)}| &>& \frac{1}{4} |w_\pi^{(0)}|, \qquad 
w_\pi^{(0)} \ < \ 0, \qquad 
w_0^{(0)}\ > \ 0. 
\end{eqnarray}

\subsection{Moduli stabilization and uplifting}
From (\ref{eq:sspgen}), the SUSY 
point is given by 
\begin{eqnarray}
e^{-\pi k T} &=& w_0/w_\pi. 
\label{eq:ekt}
\end{eqnarray}
Here we choose the parameters as 
\be
w_0 \equiv w_0^{(0)} \ > \ 0, \qquad 
w_\pi \equiv -w_\pi^{(0)} \ > \ 0, \qquad 
w_0/w_\pi \ \ll 1. 
\ee
This SUSY point has a negative vacuum energy 
$V=-3m_{3/2}^2$ where 
\begin{eqnarray}
m_{3/2}^2 &=& e^K|W|^2 
\ = \ \frac{k^3}{2}w_0^2 (1-w_0^2/w_\pi^2)^2, 
\label{eq:lsm32}
\end{eqnarray}
is the gravitino mass squared. 
The mass-square eigenvalues $m_{T \pm}^2$ 
of modulus $T$ at this point are given by 
\begin{eqnarray}
\left( m_{T+}^2, m_{T-}^2 \right) 
&=& \left( 4 m_{3/2}^2, \ 0 \right). 
\label{eq:mt2}
\end{eqnarray}
It is interesting that the stabilized value~(\ref{eq:ekt}) 
of the modulus~$T$ is the same as 
that of the KKLT model~\cite{Choi:2004sx} 
in spite of the difference between 
their effective K\"{a}hler potentials. 
Note also that the former is the exact result 
while the latter is only an approximate one that is only valid 
for large $\pi kR$. 
%
Furthermore remark that the magnitudes of the mass eigenvalues 
(\ref{eq:mt2}) are smaller than those in the KKLT model 
$m_{T \pm}^{\rm KKLT} \simeq 2\pi kR m_{3/2}$ for $\pi kR \gg 1$. 
This is because the SUSY mass contributions from the 
K\"ahler potential is comparable to those from the 
superpotential in our model, and 
they partially ($m_{T +}^2$) 
or completely ($m_{T -}^2$) cancel each other. 

Now we study the effect of the uplifting in this model. 
We uplift the AdS minimum~(\ref{eq:ssp}) by a 
sequestered vacuum energy $U$ localized at $y=0$, given by 
\begin{eqnarray}
U &=& \int d\theta^4 
(\phi \bar\phi)^n \kp \theta^2 \bar\theta^2 
\ = \ \kp e^{nK/3}. 
\end{eqnarray}
The total scalar potential is thus $V_{\rm tot}\equiv V+U$, 
and the minimum would 
be shifted as $T=T_{\rm SUSY}+\delta T$. 
We tune the constant~$\kp$ as 
\begin{eqnarray}
\kp &=& 3 e^{-nK/3} m_{3/2}^2 
\ = \ 3 \left( \frac{1-w_0^2/w_\pi^2}{2k} \right)^n m_{3/2}^2, 
\end{eqnarray}
so that $V+U=0$ at the leading order in the 
$\delta T/T_{\rm SUSY}$ expansion. Then we find the shift of 
$T$ at this Minkowski minimum as 
\begin{eqnarray}
\delta T &=& \frac{n}{2k(2+n)}, 
\end{eqnarray}
which can be small for $k \gg 1$ and $n={\cal O}(1)$. 
The SUSY breaking order parameter at this minimum is 
found as 
\begin{eqnarray}
F^T &=& \frac{n}{2k} \cdot 
\frac{w_\pi^2-w_0^2}{(2+n) w_\pi^2+n^2 w_0^2}\,m_{3/2}^2. 
\end{eqnarray}

The anomaly/modulus ratio of SUSY breaking $\alpha_{A/M}$ 
defined in Eq.~(\ref{eq:atmratio}) is calculated in this case as 
\begin{eqnarray}
\alpha_{A/M} &\simeq& 4(1+2/n). 
\end{eqnarray}
Then $\alpha_{A/M}=8$ for the typical value $n=2$ in the 
uplifting potential, and the anomaly mediation seems to 
be dominant. This should be compared with $\alpha_{A/M}=1$ in 
the KKLT model which corresponds to an asymmetric limit of our model, 
that is, $k \to 0$ in the K\"ahler potential 
keeping $k$ finite in the superpotential in (\ref{eq:ls}).

\section{Summary}
\label{sec:summary}
We studied the 4D effective theory of the 5D gauged supergravity on 
an orbifold with a universal hypermultiplet and superpotential 
terms at the orbifold fixed points. 
We analyzed a class of models obtained by gauging three independent 
isometries on the scalar manifold. 
It includes, as different limits, both the 5D heterotic 
M-theory and the SUSY RS model with an arbitrary bulk mass parameter
for the hypermultiplet. 
We have investigated the vacuum structure of such models
and the nature of moduli stabilization 
assuming perturbative superpotential terms 
at the fixed points, and discussed the uplifting of SUSY AdS vacua. 

First we analyzed the hybrid model in a case that 
the K\"{a}hler and the superpotentials in the effective action 
have analytic expressions, \ie, $(\alp,\bt)$-gauging. 
In the heterotic M-theory limit (pure $\alp$-gauging), 
the SUSY point is a saddle point of the potential, 
and the local minimum is not supersymmetric. 
The potential energies at these points are both negative, 
and thus the 4D geometry is AdS${}_4$. 
In the SUSY RS limit (pure $\bt$-gauging), on the other hand, 
the SUSY point is a local minimum with vanishing vacuum energy 
when the parameters satisfy the relation~(\ref{eq:bgsd}) 
with $c_R<0$. 
Namely this is a SUSY Minkowski vacuum. 
We have shown numerically that 
both SUSY points continuously transit to each other 
by changing the ratio $q=\bt/\alp$, 
and find that the region around the SUSY RS limit ($q\abs{c_R}\gg 1$) 
is the best candidate for the KKLT-type uplifting. 
Thus we analytically studied the uplifting 
of the SUSY AdS${}_4$ vacuum in the vicinity of the SUSY RS limit. 
For small values of $\bt$, the mirage mediation ($\alp_{A/M}=\cO(1)$) 
can be realized while the effect of the anomaly mediation is dominant 
for $\bt\simgt\cO(1)$. 
The moduli are much heavier than the gravitino in both cases. 

We also analyzed the SUSY RS model with an {\it arbitrary} bulk mass 
parameter, \ie, $(\beta,\gamma)$-gauging, and {\it generic} perturbative 
superpotential terms at the fixed points. 
If the mass parameter~$w_\vtht^{(2)}$ in the boundary superpotential 
is large enough, the matter field~$H$ is stabilized at 
$H=0$ prior to the radion $T$, and the model is reduced to 
the Luty-Sundrum model~\cite{Luty:2000ec} in the $H=0$ slice. 
Note that taking a limit $k \to 0$ 
in the K\"ahler potential while keeping finite $k$ in the 
superpotential in (\ref{eq:ls}) gives an equivalent effective theory 
to the KKLT model. 
In contrast to the KKLT model, 
the exponential terms for the modulus $T$ 
does not originate from any nonperturbative effects but from the 
warped geometry generated by the $\beta$-gauging. 
We find a SUSY AdS vacuum in this model which can be uplifted to a 
Minkowski vacuum by a sequestered SUSY breaking vacuum energy 
just like the KKLT model, 
yielding a certain anomaly/modulus ratio of the 
SUSY breaking mediation, $\alpha_{A/M} \simeq 8$. 
Note that the KKLT-type uplifting sector adopted in this paper can 
be easily extended to some dynamical SUSY breaking models like 
O'Raifeartaigh model~\cite{O'Raifeartaigh:1975pr} or 
Intriligator-Seiberg-Shih model~\cite{Intriligator:2006dd} 
as has been done in Ref.~\cite{Dudas:2006gr}. 

In this paper we have considered only a case that 
the boundary superpotentials are polynomials for 
the hypermultiplet and there are no $Z_2$-odd 
vector multiplets other than the graviphoton multiplet. 
It would be important to include 
nonperturbative effects for the study of moduli stabilization 
in more general setup. 
For the 5D heterotic M-theory, this kind of study has been done 
extensively in Refs.~\cite{Correia:2006pj,Correia:2006vf}. 
The nonperturbative effects such as the gaugino condensations generically 
depend on the gauge couplings, and thus depend on the moduli which 
determine the latter. 
The moduli dependence of the gauge couplings in the effective theory of 
5D supergravity is determined by the coefficients~$C_{IJK}$ 
in the cubic polynomial~$\cN$ in Eq.(\ref{def_cN}), 
(which is referred to as the Calabi-Yau intersection numbers 
in the 5D heterotic M-theory). 
Depending on those coefficients, we could have moduli mixings 
in the gauge couplings which play important 
role~\cite{Abe:2006xi,Abe:2005pi,Abe:2005rx} in the moduli stabilization 
with the nonperturbative effects. 
We will study these cases in our future works.

\subsection*{Acknowledgements}
H.~A. and Y.~S. are supported by the Japan Society 
for the Promotion of Science for Young Scientists 
(No.182496 and No.179241, respectively).

\appendix
\section{Isometries in on-shell description} \label{isometry}
Here we see how $SU(2,1)$ isometries are 
realized in the on-shell description of 5D supergravity. 
In order to move to the on-shell Poincar\'{e} supergravity, 
we have to fix the extraneous superconformal symmetries 
by imposing the gauge-fixing 
conditions~\cite{Kugo:2000af}-\cite{Kugo:2002js}. 
The explicit forms of these gauge-fixing conditions 
in our notation are listed 
in the appendix~A of Ref.~\cite{Abe:2006eg}. 
Since we have two compensator hypermultiplets, 
we also have to use the gauge-fixing for $U(1)_T$ 
and the equations of motion for the auxiliary fields in the $U(1)_T$ 
vector multiplet to eliminate the whole degrees of freedom 
for the compensator scalars. 
Using all the above conditions, the hyperscalars~$\vph^a$ 
($a=1,2,\ldots,6$) are expressed in terms of the physical scalar 
fields~$(S,\xi)$, which are identified with those appearing 
in Ref.~\cite{Falkowski:2000yq}, as\footnote{
These expressions are obtained from Eqs.(5.18),(5.19) and (5.26) 
in Ref.~\cite{Fujita:2001bd}. 
Here the relation between $\vph^a$ ($a=1,2,\ldots,6$) in our notation 
and the hyperscalars~$\cA_i^a$, where $i=1,2$ is the $\SUu$-index, 
in the notation of Ref.~\cite{Kugo:2000af} is $\vph^a=\cA_{i=2}^a$. 
} 
\bea
 \vph^1 \eql 0, \;\;\;\;\;
 \vph^2 = \brc{\frac{S+\bar{S}}{2(S+\bar{S}-2\abs{\xi}^2)}}^{1/2}, 
 \nonumber\\
 \vph^3 \eql \frac{\abs{1+S}}{2(S+\bar{S})^{1/2}}, \;\;\;\;\;
 \vph^4 = \vph^2\frac{\bar{\xi}(1-S)}{S+\bar{S}}, \nonumber\\
 \vph^5 \eql \vph^3\frac{1-S}{1+S}, \;\;\;\;\;
 \vph^6 = \vph^2\frac{\bar{\xi}(1+S)}{S+\bar{S}}. 
 \label{fixed_scalars}
\eea
Under the orbifold parity, $S$ and $\xi$ are even and odd, 
respectively. 
From (\ref{fixed_scalars}) we can also express $S$ and $\xi$ by 
$\vph^a$ ($a=1,2,\ldots,6$) as 
\be
 S = \frac{\vph^3-\vph^5}{\vph^3+\vph^5},  \;\;\;\;\;
 \xi = \frac{\vph^3\bar{\vph}^6-\vph^5\bar{\vph}^4}
 {\bar{\vph}^2(\vph^3+\vph^5)}.  
 \label{exp:S-xi}
\ee
Note that Eq.(\ref{exp:S-xi}) holds only in a gauge where $\vph^1=0$. 

Now we consider the $SU(2,1)$ transformations of $\vph^a$. 
They are given by 
\be
 \begin{pmatrix} \vph^{\prime 1} \\ \vph^{\prime 3} \\ \vph^{\prime 5} 
 \end{pmatrix} = e^{i\tl{\alp}\cdot T} 
 \begin{pmatrix} \vph^1 \\ \vph^3 \\ \vph^5 \end{pmatrix}, \;\;\;\;\;
 \begin{pmatrix} \vph^{\prime 2} \\ \vph^{\prime 4} \\ \vph^{\prime 6} 
 \end{pmatrix} = \brkt{e^{i\tl{\alp}\cdot T}}^* 
 \begin{pmatrix} \vph^2 \\ \vph^4 \\ \vph^6 \end{pmatrix}, 
 \label{SU21_trf}
\ee
where $\tl{\alp}\cdot T\equiv \sum_{i=1}^8\tl{\alp}_i T^i$ and 
$\tl{\alp}_i$ are the real transformation parameters. 
The $SU(2,1)$ generators~$T^i$ ($i=1,2,\ldots,8$) are given by 
\begin{eqnarray}
T^1 &=& 
\begin{pmatrix}
0 & 1 & 0 \\
1 & 0 & 0 \\
0 & 0 & 0 
\end{pmatrix}, \ 
T^2 \ = \  
\begin{pmatrix}
0 & -i & 0 \\
i & 0 & 0 \\
0 & 0 & 0 
\end{pmatrix}, \ 
T^3 \ = \
\begin{pmatrix}
1 & 0 & 0 \\
0 & -1 & 0 \\
0 & 0 & 0 
\end{pmatrix}, 
\nonumber \\
T^4 &=& 
\begin{pmatrix}
0 & 0 & 1 \\
0 & 0 & 0 \\
-1 & 0 & 0 
\end{pmatrix}, \ 
T^5 \ = \ 
\begin{pmatrix}
0 & 0 & i \\
0 & 0 & 0 \\
i & 0 & 0 
\end{pmatrix}, \ 
T^6 \ = \ 
\begin{pmatrix}
0 & 0 & 0 \\
0 & 0 & 1 \\
0 & -1 & 0 
\end{pmatrix}, 
\nonumber \\
T^7 &=& 
\begin{pmatrix}
0 & 0 & 0 \\
0 & 0 & i \\
0 & i & 0 
\end{pmatrix}, \ 
T^8 \ = \ 
\begin{pmatrix}
0 & 0 & 0 \\
0 & 1 & 0 \\
0 & 0 & -1 
\end{pmatrix}. 
\label{eq:su21gen}
\end{eqnarray}

After the transformation~(\ref{SU21_trf}), 
the compensator scalar~$\vph^{\prime 1}$ are in general nonzero. 
We can move to a gauge where $\vph^{\prime\prime 1}=0$ 
by using $\SUu$, which is part of the superconformal 
symmetries the off-shell 5D supergravity has. 
Then the other scalar components become 
\bea
 \vph^{\prime\prime 2\hat{a}-1} \eql \frac{\cos\thU}{\abs{\vph^{\prime 2}}}
 \brkt{\bar{\vph}^{\prime 2}\vph^{\prime 2\hat{a}-1}
 -\vph^{\prime 1}\bar{\vph}^{\prime 2\hat{a}}}, \;\;\;\;\;
 (\hat{a}=2,3) \nonumber\\
 \vph^{\prime\prime 2\hat{a}} \eql \frac{\cos\thU}{\abs{\vph^{\prime 2}}}
 \brkt{\vph^{\prime 1}\bar{\vph}^{\prime 2\hat{a}-1}
 +\bar{\vph}^{\prime 2}\vph^{\prime 2\hat{a}}}, \;\;\;\;\;
 (\hat{a}=1,2,3)
 \label{trfed_scalars}
\eea
where $\thU\equiv \arctan\abs{\vph^{\prime 1}/\bar{\vph}^{\prime 2}}$. 
Since $\vph^{\prime\prime 1}=0$, we can use (\ref{exp:S-xi}) 
and express the transformed physical 
scalars~$(S^{\prime\prime},\xi^{\prime\prime})$ 
in terms of the untransformed scalars~$\vph^a$, 
which are rewritten in terms of $(S,\xi)$ by the relation~(\ref{fixed_scalars}). 
Then we obtain the $SU(2,1)$ transformations of $(S,\xi)$. 
For example, 
\bea
 (S,\xi) &\to& \brkt{S-2i\tl{\alp}\xi+\tl{\alp}^2,\xi+i\tl{\alp}} \;\;\;\;\;
 \mbox{for $\tl{\alp}\cdot T=-\tl{\alp}(T^1+T^4)$} \nonumber\\
 &\to& \brkt{S+2i\tl{\alp},\xi}, \;\;\;\;\;
 \mbox{for $\tl{\alp}\cdot T=\tl{\alp}(T^6+T^8)$} \nonumber\\
 &\to& (e^{2\tl{\alp}}S,e^{\tl{\alp}}\xi), \;\;\;\;\;
 \mbox{for $\tl{\alp}\cdot T=\tl{\alp}T^7$} \nonumber\\
 &\to& (S,e^{-3i\tl{\alp}}\xi). \;\;\;\;\; 
 \mbox{for $\tl{\alp}\cdot T=\tl{\alp}(2T^3+T^8)$} 
 \label{trf:S-xi}
\eea
These transformations correspond to those of Eqs.(16),(18),(19) and (20) 
in the published version of Ref.~\cite{Falkowski:2000yq}. 
However we can show that there is no choice of $\tl{\alp}_i$ that realizes 
the transformation~(17) in Ref.~\cite{Falkowski:2000yq}, 
which we believe is their typographical error. 
On the other hand, it is easy to check that the isometries generated by 
the Killing vectors in (3.12) of Ref.~\cite{attractor} are identical 
to those derived in the above way. 
For small $\tl{\alp}_i$ ($i=1,2,\ldots,8$), general $SU(2,1)$ transformations 
of $S$ and $\xi$ are given by 
\bea
 S &\to& S+i\left\{\brkt{-\frac{\tl{\alp}_3}{2}+\tl{\alp}_6+\tl{\alp}_8}
 -2i\tl{\alp}_7 S+\brkt{\bar{\alp}_{12}+\alp_{45}}\xi 
 \right. \nonumber\\
 &&\hspace{15mm} \left. 
 +\brkt{\bar{\alp}_{12}-\alp_{45}}S\xi
 +\brkt{\frac{\tl{\alp}_3}{2}+\tl{\alp}_6-\tl{\alp}_8}S^2
 \right\}+\cO(\tl{\alp}^2), \nonumber\\
 \xi &\to& \xi+i\left\{-\frac{\alp_{12}+\bar{\alp}_{45}}{2}
 +\frac{\alp_{12}-\bar{\alp}_{45}}{2}S
 -\brkt{\frac{3}{2}\tl{\alp}_3+i\tl{\alp}_7}\xi \right. \nonumber\\
 && \hspace{15mm} \left. 
 +\brkt{\frac{\tl{\alp}_3}{2}+\tl{\alp}_6-\tl{\alp}_8}S\xi
 +\brkt{\bar{\alp}_{12}-\alp_{45}}\xi^2\right\}+\cO(\tl{\alp}^2), 
\eea
where $\alp_{12}\equiv\tl{\alp}_1+i\tl{\alp}_2$ and 
$\alp_{45}\equiv\tl{\alp}_4+i\tl{\alp}_5$.

\end{document}